\documentclass[conference]{IEEEtran}
\usepackage{cite}

\usepackage{siunitx}

\ifCLASSINFOpdf
\usepackage{graphicx}
 \graphicspath{{../pdf/}{../jpeg/}}
\DeclareGraphicsExtensions{.pdf,.jpeg,.png,.eps}
\else
\fi
\usepackage{amsmath}
\usepackage{amssymb}
\usepackage{mathtools}

\usepackage{algorithmic}
\usepackage{algorithm}
\newsavebox{\ieeealgbox}

\usepackage[tight,footnotesize]{subfigure}
\usepackage[acronym]{glossaries}
\newacronym{mimo}{MIMO}{multiple-input multiple-output}
\newacronym{v2v}{V2V}{vehicle-to-vehicle}
\newacronym{gscm}{GSCM}{geometric-based stochastic channel model}
\newacronym{dmc}{DMC}{diffuse multipath components}
\newacronym[plural = SPs, firstplural=specular paths (SPs)]{sp}{SP}{specular paths}
\newacronym{ekf}{EKF}{extended Kalman filter}
\newacronym{sage}{SAGE}{space-alternating generalized expectation-maximization}
\newacronym{io}{IO}{interaction objects}
\newacronym{mle}{MLE}{maximum likelihood estimator}
\newacronym{blue}{BLUE}{best linear unbiased estimator}
\newacronym{tx}{TX}{transmitter}
\newacronym{rx}{RX}{receiver}
\newacronym{ddtf}{DDTF}{double directional transfer function}
\newacronym[plural = APDPs, firstplural=average power delay profiles (APDPs)]{apdp}{APDP}{average power delay profile}
\newacronym{tdoa}{TDoA}{time delay of arrival}
\newacronym{dod}{DoD}{direction of departure}
\newacronym{doa}{DoA}{direction of arrival}
\newacronym{3d}{3D}{3-dimensional}
\newacronym{4d}{4D}{4-dimensional}
\newacronym{cdf}{CDF}{cumulative distribution function}
\newacronym{hrpe}{HRPE}{high resolution parameter estimation}
\newacronym[plural = MPCs, firstplural=multipath components (MPCs)]{mpc}{MPC}{multipath components}
\newacronym{fim}{FIM}{Fisher Information Matrix}
\newacronym{em}{EM}{Expectation Maximization}
\newacronym{tdm}{TDM}{time-division multiplex}
\newacronym{snr}{SNR}{signal-to-noise ratio}
\newacronym{pdp}{PDP}{power delay profile}
\newacronym{ofdm}{OFDM}{orthogonal frequency division multiplexing}
\newacronym{lo}{LO}{local oscillator}
\newacronym{pps}{PPS}{pulse per second}
\newacronym{pa}{PA}{power amplifier}
\newacronym{los}{LoS}{line-of-sight}
\newacronym{gps}{GPS}{global positioning system}
\newacronym{siso}{SISO}{single-input single-output}
\newacronym{uca}{UCA}{uniform circular array}
\newacronym{eadf}{EADF}{effective aperture distribution function}
\newacronym{aps}{APS}{angular power spectrum}
\newacronym{t2t}{T2T}{truck-to-truck}
\newacronym{dpr}{dPR}{diffuse power ratio}
\newacronym{ifft}{IFFT}{inverse fourier transform}
\newacronym{cmd}{CMD}{correlation matrix distance}
\newacronym{simo}{SIMO}{single input multiple output}
\newacronym{nlos}{NLoS}{non-line-of-sight}
\newacronym{pl}{PL}{pathloss}

\newcommand{\RNum}[1]{\uppercase\expandafter{\romannumeral #1\relax}}
\newcommand{\tx}[1]{\text{#1}}

\usepackage[utf8]{inputenc}
\usepackage[table]{xcolor}
\usepackage{booktabs}
\usepackage{gensymb}
\usepackage{siunitx}

\begin{document}
%
\title{Stationarity Region of Mm-Wave Channel Based on Outdoor Microcellular Measurements at 28 GHz}

\author{\IEEEauthorblockN{R. Wang$^1$, \textit{Student Member, IEEE}, C. U. Bas$^1$, \textit{Student Member, IEEE},\\ S. Sangodoyin$^1$, \textit{Student Member, IEEE}, S. Hur$^2$, \textit{Member, IEEE},\\ J. Park$^2$, \textit{Member, IEEE}, J. Zhang$^3$, \textit{Fellow, IEEE}, A. F. Molisch$^1$, \textit{Fellow, IEEE}}
\IEEEauthorblockA{$^1$University of Southern California, Los Angeles, CA USA,\\ $^2$ Samsung Electronics, Suwon, Korea, \\ $^3$ Samsung Research America, Richardson, TX, USA}}


%


\maketitle

\begin{abstract}
The stationarity region, i.e., the area in which the {\em statistics} of a propagation channel remain constant, is an important measure
of the propagation channel, and essential for efficient system design. This paper presents what is to our knowledge the first extensive measurement campaign
for measuring the stationarity region of MIMO mm-wave channels. Using a novel 28 GHz phased-array sounder with high phase stability, we present results 
in an urban microcell LoS, and LOS to NLOS transition region scenario, for the stationarity region of shadowing, power delay profile, and the angular power spectrum. A comparison to results at cm-waves shows considerably reduced stationarity region size, which has an important impact on system design. 

\end{abstract}


%
\IEEEpeerreviewmaketitle

\section{Introduction}

One of the most important ways to satisfy the constantly increasing demand for wireless data is making available more spectrum for cellular and WLAN communications. 
An especially promising frequency range is the millimeter-wave (mm-wave) band, due to the large swaths of currently fallow spectrum \cite{andrews2014will}. In the USA, the frequency regulator
(Federal Communications Commission, FCC) has recently made available more than \SI{15}{GHz} of bandwidth for unlicensed or licensed operation, and other countries, as well
as the World Radio Conference, are expected to follow suit. 
Furthermore, recent progress in semiconductor technology and adaptive beamforming 
has made possible the fabrication of mm-wave hardware at reasonable prices \cite{rangan2014millimeter}. For all these reasons, the 5G standard by 3GPP (the international standardization body for cellular communications) as well 
as the IEEE 802.11 standard for WLANs prominently feature mm-wave communications.  

In order to design and evaluate wireless systems, a thorough understanding of the propagation channel is required \cite{molisch2011wireless}. In this context, it is important to note that mm-wave channels show fundamentally different behavior compared to channels below \SI{6}{GHz} carrier frequency, which have been mostly investigated up to now. These differences are an almost complete absence of diffraction as effective propagation process, and the relatively greater roughness (in units of wavelength) of objects the waves interact with \cite{shafi20175g}. For these reasons, a number of propagation measurements have been done in both indoor and outdoor environments at mm-waves over the past 30 years (see \cite{smulders2009statistical, haneda2015channel, shafi20175g} for a review and further references. Recent work has particularly concentrated on directional channel characteristics, i.e., the channel behavior as seen by directional antennas since use of directional antennas is essential at mm-wave frequencies as the antenna gain is required to compensate for the higher free-space \gls{pl}. Standardized channel models for IEEE 802.11ad \cite{Maltsev_et_al_2010} and 3GPP SCM \cite{haneda20165g} have been derived. Small-scale fading statistics have been analyzed, e.g., in \cite{samimi201628}.

However, despite the existence of some measurements, many gaps in our understanding of propagation channels remain. One of the most important gaps is the {\em dynamic} behavior of the channels. Our results attempt to answer the question about the {\em stationarity region} of the channel, i.e., the area over which the {\em channel statistics} stay approximately constant. This is so important for mm-wave systems because on one hand due to the mobility of users, fixed-orientation horn or parabolic antennas cannot be used in cellular systems and on the other hand, adaptive beamforming is often based on the channel statistics (and not the instantaneous fading) in order to reduce channel training overhead and hardware requirements \cite{roh2014millimeter}. Furthermore, the area in which the \gls{pdp} / delay spread stays constant is critical for efficient pilot design, while gain control and similar parameters may depend on the shadowing state. For all these reasons, measurements of the stationarity region size is essential. At cm-wave frequencies, numerous investigations of stationarity region size have been done, and metrics have been established to measure the stationarity region size of shadowing, PDP, and \gls{cmd} \cite{matz2005non, paier2008non, he2015characterization} and references therein. 

Yet, at mm-wave frequencies, hardly any measurements exist, with the exception of the few cases discussed below. For example, the corresponding parameters in the 3GPP channel model are not based on mesaurements at mm-wave frequencies. Ref. \cite{liya2016characterization} investigates the lifetime of multipath components in a street canyon, as well as their change in power and direction. However, the measurements were only done in a \gls{los} scenario in a single street canyon, measured at 7 locations separated by \SI{5}{m} each, and furthermore are \gls{simo}, not \gls{mimo}. Also, measurements were evaluated with respect to the persistence and angle change of individual multipath components, which are only indirectly related to stationarity regions. Some recent work \cite{maccartney2017flexible,sun2017millimeter,rappaport2017small}, done in parallel to our investigations, investigates the nonstationary statistics of mm-wave channel, which include the diffraction measurement and the transition between \gls{los} and \gls{nlos}, but is based on experiments at \SI{73}{GHz}, in contrast to our investigations at \SI{28}{GHz}. Results are presented for the change in the pathloss along a track, or in a cluster, with measurement locations spaced \SI{5}{m} apart. However, to our knowledge there are no measurements in the literature that are based on a dense spatial sampling; nor are there measurements of stationarity regions for shadowing, PDP, or angular spectra. In this paper, we aim to fill this gap. 

The main contributions of this paper are as follows:
\begin{itemize}
\item We present the measurement setup and procedure for {\em MIMO} measurements on a {\em densely sampled} route, using a novel phase array based channel sounder with high phase stability. Our results are based on more than 29 million recorded impulse responses. 
\item For measurement routes covering \gls{los} and the transition between \gls{los} and \gls{nlos}, we present stationarity regions of channel statistics such as \gls{pdp} and shadow fading, as well as the evolution of angular spectra.  
\end{itemize}
The results thus serve as important guidelines for mm-wave system design. 

The remainder of the paper is organized as follows: Section II describes our channel sounder and the measurement environment and procedure. Section III covers the evaluation procedure of the data, and presents the results. A summary and conclusions in Section IV wrap up the paper.

\section{Measurement Campaign}
\label{sect:Meas}

In this campaign, we used a real-time, phased-array, wide-band mm-wave channel sounder that is capable of beam-forming at both the \gls{rx} and the \gls{tx} \cite{bas_2017_realtime}. Both the \gls{tx} and the \gls{rx} are equipped with phased antenna arrays capable of forming beams which can be electronically steered with $5^{\circ}$ resolution in the range of $[-45^{\circ}, 45^{\circ}]$ in azimuth. Compared to the setups with rotating horn antennas or virtual antenna arrays, the electronically switched beams decreases the measurement time for one \gls{rx}-\gls{tx} location from hours to milliseconds. During this campaign, we measured for 361 total beam pairs in \SI{14.44}{ms} per location, with a \gls{siso} averaging factor of 10 allowing a measurable path-loss of \SI{169}{dB}. Thanks to the short measurement time, unlike the rotating horn antenna sounders, we can perform double directional measurements on continuous routes as the \gls{rx} or \gls{tx} moves. By using GPS-disciplined Rubidium frequency references, we were able to achieve both short-time and long-time phase stability. Combined with the short measurement time this limits the phase drift between \gls{tx} and \gls{rx} even when they are physically separated and have no cabled connection for synchronization as in this case. Consequently, all \gls{tx} and \gls{rx} beams were measured phase-coherently.

The sounding signal has 801 tones equally spaced over \SI{400}{MHz} total bandwidth. This provides a \SI{2.5}{ns} delay resolution with $\SI{2}{\mu s}$ measurable excess delay. To better utilize the dynamic range of the sounder, the phases of individual tones are manipulated as suggested in \cite{Friese1997multitone} resulting in a peak to average power ratio (PAPR) of \SI{0.4}{dB}. The main specifications for the sounder  and sounding signal are listed in Table \ref{specs}. With the flexible FPGA control interface, many of these operation parameters such as; the number of and the order of \gls{tx}-\gls{rx} beam pairs, the \gls{siso} duration and the number of SISO repetition per beam pair, the measurable excess delay and the MIMO repetition rate can be modified on a per campaign basis. For example, for measurements in highly dynamic environments, we can use only the odd numbered beams (without missing any multi-path components, since the beam width is larger than twice the beam step) with no \gls{siso} averaging, and lower the measurement time for a MIMO sweep to $\SI{400}{\mu s}$. Further details of the sounder and the validation measurements can be found in \cite{bas_2017_realtime}.

\begin{table}[tbp]\centering
  \caption{Sounder specifications}
  \renewcommand{\arraystretch}{1.1}
\begin{tabular}{l|c}
    \hline
    \multicolumn{2}{c}{Hardware Specifications} \\ \hline \hline
    Center Frequency & \SI{27.85}{GHz}\\
    Instantaneous Bandwidth & \SI{400}{MHz}\\
    Antenna array size & 8 by 2 (for both \gls{tx} and \gls{rx}) \\
    Horizontal beam steering & $-45^\circ$ to $45^\circ$ \\
    Horizontal 3dB beam width & $12^\circ$\\
    Horizontal steering steps & $5^\circ$\\
    Beam switching speed & $\SI{2}{\mu s}$ \\
    \gls{tx} EIRP & 57 dBm \\
    \gls{rx} noise figure & $\le$ 5 dB \\ 
    ADC/AWG resolution & 10/15-bit \\
    Data streaming speed & \SI{700}{MBps} \\ \hline
    \multicolumn{2}{c}{Sounding Waveform Specifications} \\ \hline \hline
    Waveform duration & $\SI{2}{\mu s}$ \\
    Repetition per beam pair & 10 \\
    Number of tones & 801 \\
    Tone spacing & \SI{500}{kHz} \\
    PAPR & \SI{0.4}{dB} \\ 
    Total sweep time & \SI{14.44}{ms}\\ 
    MIMO repetition rate & \SI{5}{Hz}\\ 
    \hline      
  \end{tabular} \label{specs}
\end{table}

 \begin{figure*}[!t]
   \centering
  \includegraphics[width=0.8\textwidth, clip = true]{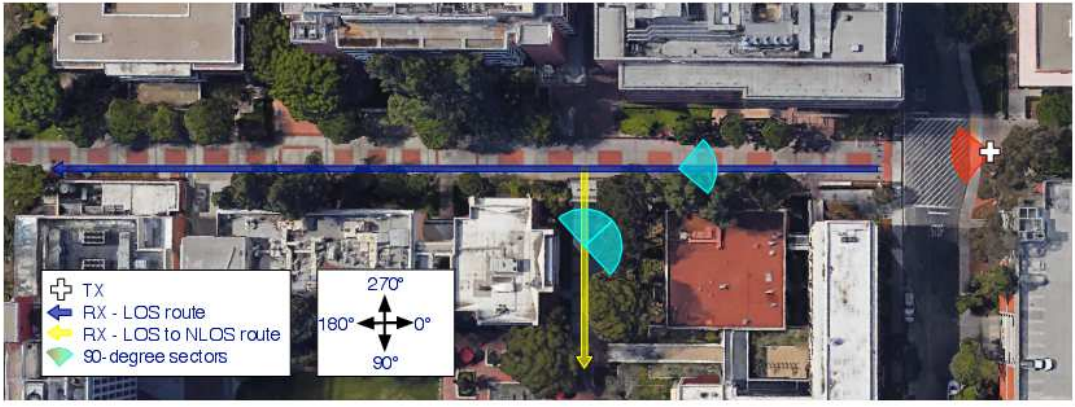}\caption{Measurement Environment}\label{Fig:meas_locs}
\end{figure*}

The measurements were performed on the University Park Campus, University of Southern California, in Los Angeles, CA, USA, in an area that resembles a typical urban environment. The \gls{tx} was placed on a scissor lift at the height of \SI{5}{m} while the \gls{rx} was on a mobile cart with the antenna height of \SI{1.8}{m}. We captured one MIMO snapshot every \SI{200}{ms} as we moved the \gls{rx} with an approximate speed of \SI{0.15}{m/s} on straight lines resulting in a spatial sampling with an average rate of \SI{3}{cm} per sample. As the \gls{rx} moves, a laser distance meter records the distance to a reference object placed at the start of the route. 
 
We focused on two scenarios; \gls{los} and \gls{los} to \gls{nlos} transition. For all measurements we performed $[-45^{\circ}, 45^{\circ}]$ azimuth sweeps at both the \gls{rx} and the \gls{tx}, while the bore-sights of the arrays facing towards to possible paths of propagation. The \gls{los} measurements cover a \SI{160}{m} route, and in this case $90^{\circ}$ \gls{rx} sector was facing $0^{\circ}$. The \gls{los} to \gls{nlos} transition measurements were captured on a \SI{40}{m} route which starts from \SI{53}{m} into the \gls{los} route. For the transition, we did two runs to include both when the \gls{rx} sector is oriented at $0^{\circ}$ (i.e. same orientation with \gls{los}) and $270^{\circ}$ (i.e. facing towards the \gls{los} route). All routes along with \gls{rx}/\gls{tx} orientations are depicted in Figure \ref{Fig:meas_locs}.

\section{Evaluation}
\subsection{Data postprocessing}
The directional \gls{pdp} is estimated from the measured transfer function $H$ according to
\begin{equation}
  P_h(m,n,x,\tau) = \vert \mathcal{F}^\tx{-1}\big\{ w(f)H(m,n,x,f) \big\} \vert^2,
\end{equation}
where $m = 1,2,...,M_t$ is the \gls{tx} beam index, $n=1,2,...,M_r$ is the \gls{rx} beam index, and $x$ is the displacement of \gls{rx} from the origin of the continuous route. A gain-corrected Hanning window $w(f)$ is applied on the transfer function $H$ before taking \gls{ifft}. We have also compensated the transfer function based on the system calibration response.

Thanks to the precise laser distance meter, we can compute the average \gls{pdp} over small-scale fading in a distance window with length $d_0 = \SI{40}{cm}$, which is approximately $\SI{40}{\lambda}$ at \SI{28}{GHz}. The $i$th average \gls{pdp} is given by
\begin{equation}
  \bar{P}_h(m,n,i,\tau) = \frac{1}{N_i} \underset{x \in D_i}{\sum} P_h(m,n,x,\tau), \label{Eq:ave_PDP}
\end{equation}
where $D_i = \big[(i-1)d_0,i \cdot d_0 \big)$, $N_i$ is the number of \gls{mimo} snapshots taken during this distance window $D_i$.

Similar to the spirit of \cite{hur2014synchronous,bas_2017_realtime}, the omni-directional \footnote{Strictly speaking the result can only be applied to the $90^\circ$ openings at both \gls{tx} and \gls{rx}} \gls{pdp} is computed as
\begin{equation}
  \bar{P}_{h,\tx{max}}(i,\tau) = \underset{m,n}{\tx{max}}\;  \bar{P}_h(m,n,i,\tau). 
\end{equation}
As a result the received power can be computed by
\begin{equation}
  P_{RX}(d_i) = \underset{\tau \in S_{\tau,i}}{\sum} \bar{P}_{h,\tx{max}}(i,\tau),
\end{equation}
where $S_{\tau,i}$ is the set of delay bins whose power is \SI{3}{dB} higher than the noise floor in $\bar{P}_{h,\tx{max}}(i,\tau)$. The propagation loss in dB is then given as
\begin{equation}
  PL(d_i) = P_t - 10\tx{log}_{10}(P_{RX}(d_i)),
\end{equation}
where $P_t$ is the transmit power and kept at constant during the expriments.

\subsection{Pathloss fitting}
\begin{figure}[!t]
  \centering
  \includegraphics[width = 0.85\columnwidth, clip = true]{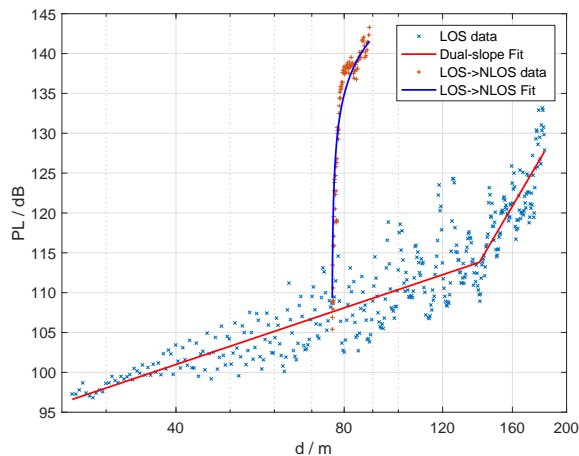}
  \caption{The fitting of pathloss data, \gls{los} data based on a dual-slope model given by Eq. (\ref{Eq:PL_los_dualslope}), \gls{los}$\rightarrow$\gls{nlos} data based on the street-by-street PL model in \cite{7481421}}
  \label{fig:PL_fit_dualSlope}
\end{figure}

Fig. \ref{fig:PL_fit_dualSlope} presents the pathloss fitting result for the \gls{los} route. Due to the foliage effect, the received power for the tail of the \gls{los} route is significantly attenuated.\footnote{Note that this is {\em not} a breakpoint due to the classical ground reflection model - that breakpoint is at \SI{10}{km}.} The results demonstrate the importance of including foliage effect in mm-wave system simulations. We fit the propagation loss data to a dual-slope model to acquire the pathloss exponents $(n_1,n_2)$, cut-off distance $d_c$ and the intercept offset $b$ The fitting model is given as
\begin{equation}
  \overline{PL}_\tx{los}(d) = \begin{cases}
     10n_1 \tx{log}_{10}(d) + b, &\text{if }d < d_c \\
     10n_1 \tx{log}_{10}(d_c) + 10n_2 \tx{log}_{10}(\frac{d}{d_c}) + b, &\text{otherwise}
   \end{cases}
   \label{Eq:PL_los_dualslope}
\end{equation}
where $d$ is the distance between \gls{tx} and \gls{rx}. The fitted parameters are listed in Tab. \ref{Tab:PL_Params}. 
\begin{table}[!t]
  \centering
  \caption{Parameters of the pathloss model for the \gls{los} and \gls{los} to \gls{nlos} transition scenarios} 
  \label{Tab:PL_Params}
  \begin{tabular}{|l|c|}
    \toprule
    Parameters & Values \\
    \midrule
    $n_1$ & 2.36 \\
    $n_2$ & 11.93 \\
    $b$   & 63.18 \\
    $d_c$ & 139.64\\
    $n_{tr}$ & 2.59 \\
    $\Delta$ & 0 \\
    \bottomrule
  \end{tabular}
\end{table}

\begin{figure}[!t]
  \centering
  \includegraphics[width = 0.85\columnwidth, clip = true]{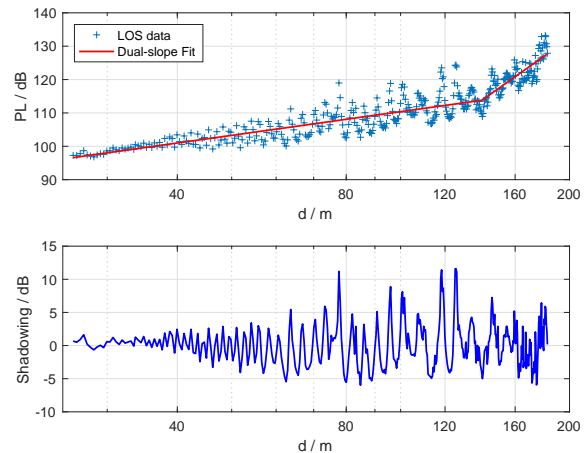}
  \caption{The shadowing realization extracted from the \gls{los} route}
  \label{fig:Sh_los}
\end{figure}

The realization of the shadowing process is the residual component of this fitting process, thus determined by
\begin{equation}
  Sh(d_i) = PL(d_i) - \overline{PL}_\tx{los}(d_i)
\end{equation}
Let us investigate the shadowing process of measurement points that experience an unobstructed \gls{los}, i.e. $d < d_c$. Assuming the shadowing process is stationary during this part of the route, the autocorrelation function of the shadowing process can be evaluated according to
\begin{equation}
  A_{Sh}(\Delta d) = \frac{\mathbb{E}_d \{ Sh(d)Sh(d+\Delta d)\}}{\mathbb{E}_d \{Sh^2(d)\}},
\end{equation}
where $\mathbb{E}\{\}$ is the expectation operator. The standard deviation of shadowing is \SI{3.11}{dB}, and the autocorrelation distance is \SI{1.2}{m} when we set the decorrelation threshold as $1/e$. The relatively short shadowing correlation distance can be explained by the oscillation of propagation loss around the mean path loss, which is illustrated in Fig. \ref{fig:Sh_los}. This phenomenon is also observed in the typical two-path channel \cite{molisch2011wireless}. While interference between multipath components is normally not counted as ``shadowing'', it could be eliminated here only by using a larger averaging window, which would entail the danger of averaging out some actual shadowing as well. This is a demonstration of the general difficulty of separating small-scale and large-scale fading, an effect that occurs at all frequencies \cite{wyne2009statistical}.

We also analyze the propagation loss for the \gls{los}-to-\gls{nlos} transition route, where \gls{rx} has two orientations. Let us denote them as $PL_{tr,1}(y)$ and $PL_{tr,2}(y)$, where the subscript 1 and 2 is used to differentiate the two routes measured for \gls{los} to \gls{nlos} transition. We analyze the smaller propagation loss at the same location, which is defined as
\begin{equation}
  PL_{tr}(y) = \underset{i\in\{1,2\}}{\tx{min}} \; PL_{tr,i}(y)
\end{equation}
We observe a large and rapid increase of propagation loss when \gls{rx} moves from \gls{los} to \gls{nlos}. The path loss increase is about \SI{30}{dB}, which is close to the \SI{25}{dB} reported in \cite{rappaport2017small}. Following the method in \cite{7481421}, we propose to fit this \gls{los} to \gls{nlos} transition part according to the model given by
\begin{equation}
  PL_{tr}(y) = 10n_{tr}\tx{log}_{10}(y) + \Delta + \overline{PL}_\tx{los}(d_1) 
\end{equation}
Here $y$ is the local displacement of \gls{rx} from the starting point of the transition route, $\Delta$ denotes the corner loss proposed in \cite{7481421} and is constrained to be nonnegative during the data fitting, and finally $d_1$ the distance between  the route starting point and \gls{tx}. The corresponding measured data and the fitted line is presented in Fig. \ref{fig:PL_fit_dualSlope}, and the parameters are given in Tab. \ref{Tab:PL_Params}. The standard deviation of the shadowing process is \SI{3.4}{dB} and the correlation distance is \SI{4.8}{m}.

\subsection{Correlation of PDP}
\begin{figure}[!t]
  \centering
  \includegraphics[width = 0.85\columnwidth, clip = true]{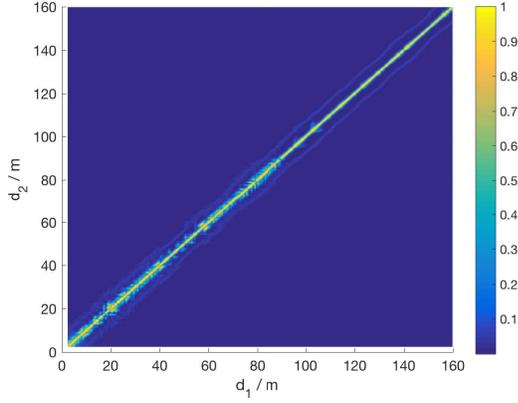}
  \caption{The collinearity function of PDP for the \gls{los} route}
  \label{fig:PDP_corr_los}
\end{figure}

\begin{figure}[!t]
  \centering
  \includegraphics[width = 0.85\columnwidth, clip = true]{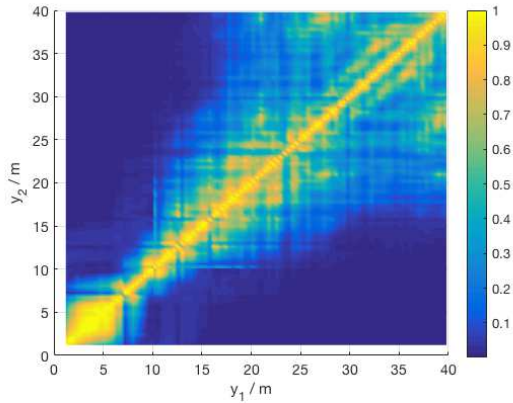}
  \caption{The collinearity function of PDP for the \gls{los} to \gls{nlos} route 2, when \gls{rx} faces \gls{tx}}
  \label{fig:PDP_corr_los_nlos}
\end{figure}

We also evaluate the similarity of the omni-directional PDP $\bar{P}_{h,\tx{max}}(i,\tau)$ measured along the continous \gls{rx} route, according to
\begin{equation}
  X_\tx{pdp}(i,j) = \frac{\int_\tau \bar{P}_{h,\tx{max}}(i,\tau) \bar{P}_{h,\tx{max}}(j,\tau)\, \tx{d}\tau }{\sqrt{\int_\tau \vert \bar{P}_{h,\tx{max}}(i,\tau) \vert^2 \,\tx{d}\tau} \sqrt{\int_\tau \vert \bar{P}_{h,\tx{max}}(j,\tau) \vert^2\, \tx{d}\tau}}
\end{equation}
We can easily verify that $X_\tx{pdp}(i,i) = 1, \, \forall i$. The analysis is similar to the one presented in \cite{paier2008non} for the collinearity of \gls{pdp} or local scattering function (LSF) in vehicle-to-vehicle propagation channels at \SI{5.8}{GHz}. Fig. \ref{fig:PDP_corr_los} presents the collinearity function of the \gls{pdp} for the \gls{los} route, where the same function is evaluated for \gls{los} to \gls{nlos} transition route 2 in Fig. \ref{fig:PDP_corr_los_nlos}. If we set the threshold of de-correlation for $X_\tx{pdp}(i,j)$ as 0.9, the average correlation distance is about \SI{0.9}{m} for the LOS scenario, but it can be as high as \SI{4}{m} at the beginning section of the transition route and drops to about \SI{1.26}{m} afterwards. These are comparatively smaller than the values presented in \cite{paier2008non}. For the \gls{los} case, the short correlation distance is a consequence of the high delay resolution, as the \gls{los} moves from one resolvable delay bin to another with a relatively short movement of the RX. This has important consequences for the design of loop bandwidth in timing acquisition/tracking. 

\subsection{Analysis with time-varying APS}
\begin{figure}[!t]
  \centering
  \includegraphics[width = 0.85\columnwidth, clip = true]{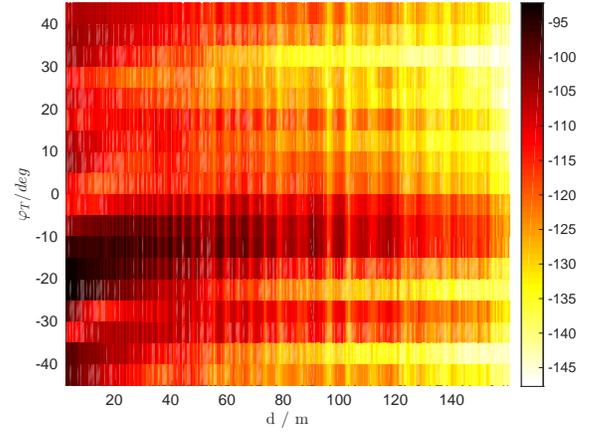}
  \caption{The \gls{aps} for \gls{tx} on the \gls{los} route}
  \label{fig:APS_tx}
\end{figure}

\begin{figure}[!t]
  \centering
  \includegraphics[width = 0.85\columnwidth, clip = true]{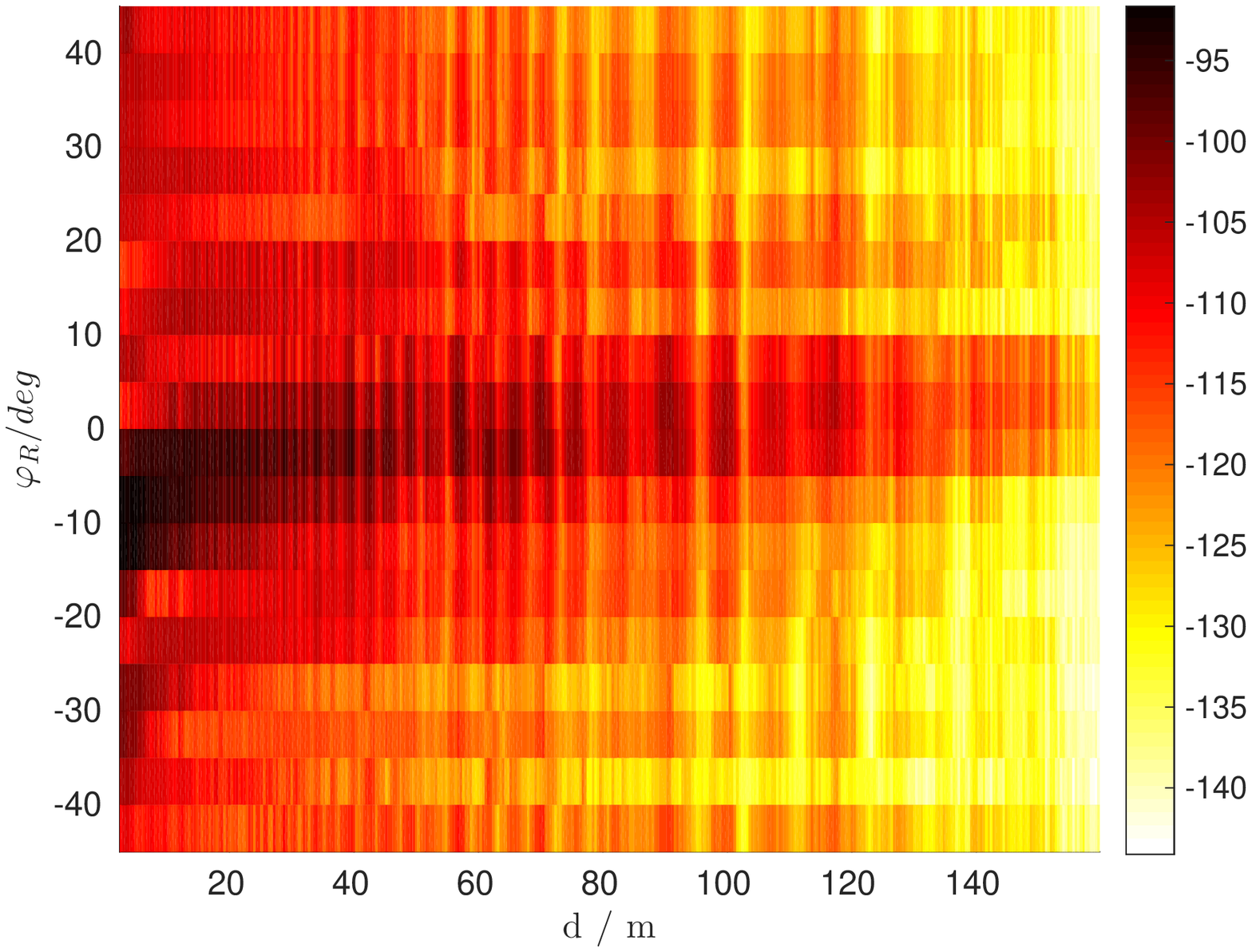}
  \caption{The \gls{aps} for \gls{rx} on the \gls{los} route}
  \label{fig:APS_rx}
\end{figure}

\begin{figure}[!t]
  \centering
  \includegraphics[width = 0.85\columnwidth, clip = true]{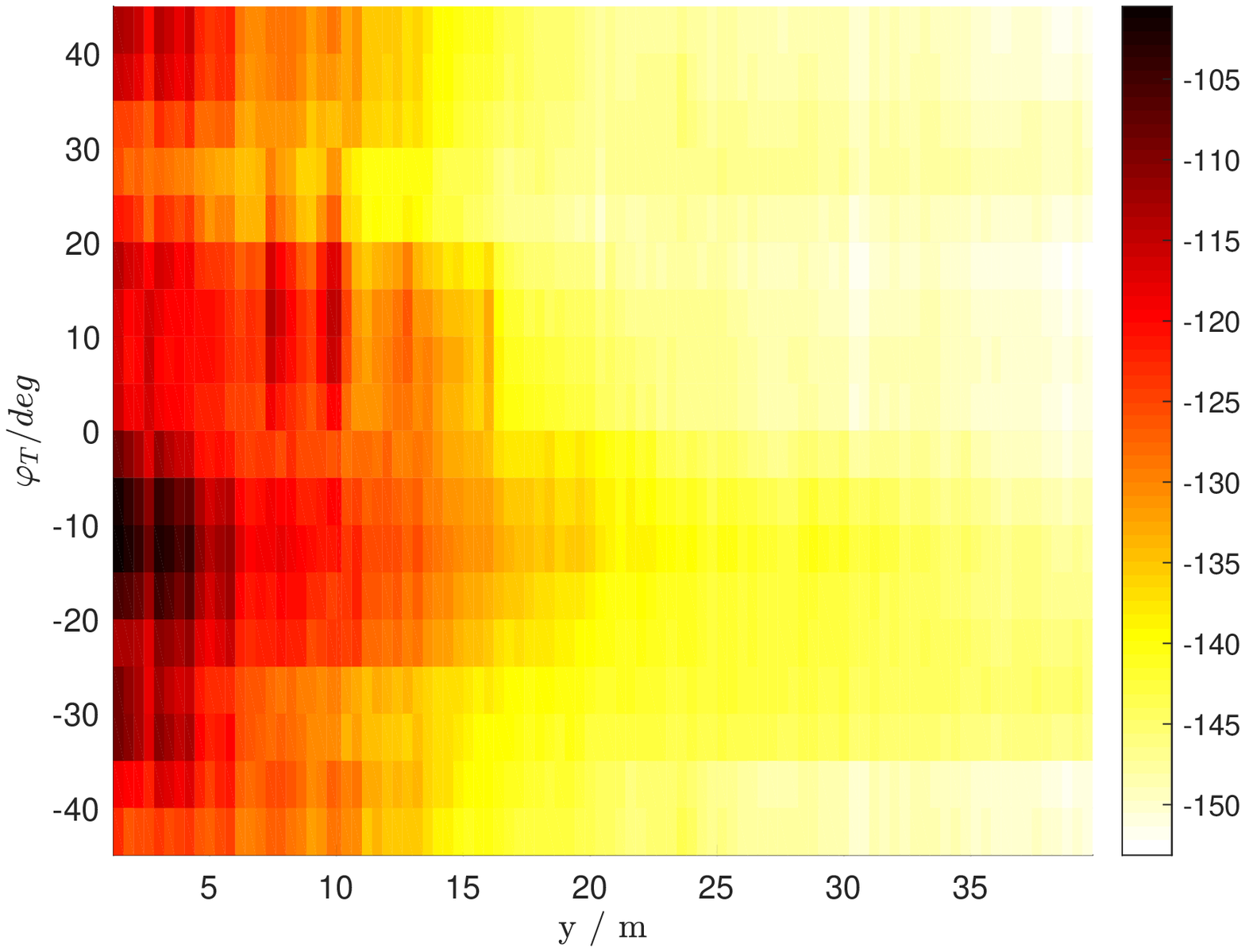}
  \caption{The \gls{aps} for \gls{tx} on the \gls{los} to \gls{nlos} transition route, when \gls{rx} faces $0^\circ$}
  \label{fig:APS_tx_los_nlos_2}
\end{figure}

\begin{figure}[!t]
  \centering
  \includegraphics[width = 0.85\columnwidth, clip = true]{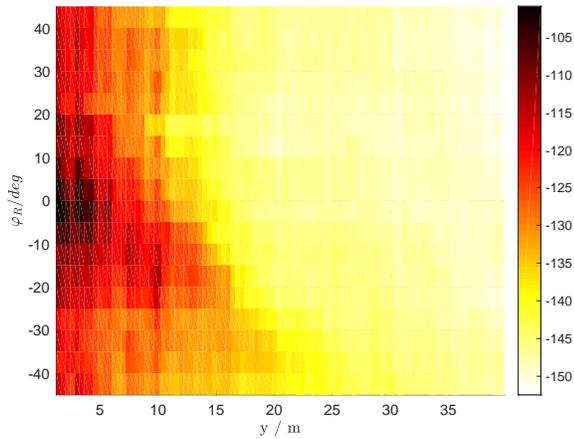}
  \caption{The \gls{aps} for \gls{rx} on the \gls{los} to \gls{nlos} transition route, when \gls{rx} faces $0^\circ$}
  \label{fig:APS_rx_los_nlos_2}
\end{figure}

We also analyze the time-varying \gls{aps} for both \gls{tx} and \gls{rx} along the two routes. The results provide important insight into the directional information about dominant \glspl{mpc} when the \gls{rx} moves along a continuous track. Similar to the definition in \cite{bas_2017_realtime}, we can compute the \gls{aps} based on the average PDP from Eq. (\ref{Eq:ave_PDP}), which is given by
\begin{align}
  P(\varphi_T,d_i) &= \sum_{n,\tau \in S_\tau^{m,n,i}} \bar{P}_h(m,n,i,\tau) \\
  P(\varphi_R,d_i) &= \sum_{m,\tau \in S_\tau^{m,n,i}} \bar{P}_h(m,n,i,\tau)
\end{align}
Here $\varphi_T$ is the azimuth \gls{dod} and ranges between -$45^\circ$ and $45^\circ$ with a $5^\circ$ angle increment. Similarly we have $\varphi_R$ for the azimuth \gls{doa}. The directions of $0^\circ$ in both cases are aligned with the orientations for \gls{tx} and \gls{rx} orientations. More details are presented in Section \ref{sect:Meas}. Figs. \ref{fig:APS_tx} and \ref{fig:APS_rx} are for the \gls{los} route, where we can match the dominant direction with \gls{los}. Figs. \ref{fig:APS_tx_los_nlos_2} and \ref{fig:APS_rx_los_nlos_2} are for the \gls{los} to \gls{nlos} transition route, where we clearly see the dominant paths shift towards -$40^\circ$ when \gls{rx} moves into \gls{nlos} region. They are most likely reflected and diffracted paths around the street corner. This adaptation of the beam direction is critical for beam tracking, and depending on the speed of the \gls{tx}/\gls{rx}, relatively quick determination of new directions would have to be taken into account. These changes occur at {\em both} the \gls{tx} and the \gls{rx}.

\section{Conclusion}
In this paper we present, to our knowledge, the first extensive measurement results on the stationarity region of MIMO mm-wave channels, which are based on over 20 million measured channel impulse responses. We have found that the foliage effect can significantly alter the \gls{pl} exponent even in the \gls{los} scenario, which leads the channel nonstationarity captured by our proposed dual-slope \gls{pl} model. Meanwhile the propagation loss increases rapidly, about \SI{30}{dB}, during the transition from \gls{los} to \gls{nlos} scenario. We propose to model it with a street-by-street \gls{pl} model. The autocorrelation distance of shadowing is \SI{1.2}{m} in \gls{los} route, although this value might be affected by the signal variation observed in the two-path channel. It rises up to \SI{4.8}{m} for the \gls{los} to \gls{nlos} transition route. The average correlation distance, computed based on the similarity of \gls{pdp}, is \SI{0.9}{m} for the \gls{los} route. It can reach as high as \SI{4}{m} at the beginning section of the transition route and drops to about \SI{1.26}{m} afterwards. The analysis based on the correlation of shadowing and \gls{pdp} has suggested that mm-wave channels exhibit a smaller stationarity region compared with cm-wave channels. 


\section*{Acknowledgment}
Part of this work was supported by grants from the National Science Foundation. The authors would like to thank Dr. Dimitris Psychoudakis, Thomas Henige, Robert Monroe for their contribution in the development of the channel sounder.



%

\bibliographystyle{IEEEtran}
\bibliography{mmWave_SR_main_CR}

\end{document}